# Deterministic Exciton Confinement in 2D Semiconductors via Local Dielectric Engineering for Scalable Quantum Light Sources

R. Itzhak[1,2], A. Hayat[1], and I. Goykhman[2*]

[1] Micro Nanoelectronics Research Center, Department of Electrical and Computer Engineering, Technion, Haifa, Israel.
[2] Institute of Applied Physics, The Faculty of Science and The Center for Nanoscience and Nanotechnology, The Hebrew University of Jerusalem, Jerusalem 91904, Israel.

*Contact author: ilya.goykhman@mail.huji.ac.il

**ABSTRACT**. Single-photon emitters (SPEs) are essential building blocks for quantum communication and photonic quantum technologies. However, realizing scalable, on-chip SPEs in a CMOS-compatible platform remains a major challenge. Here, we propose and theoretically demonstrate a novel approach for exciton confinement in two-dimensional (2D) semiconductors via local dielectric engineering. By introducing a high-dielectric-constant (high-$k$) nanopillar above or beneath the 2D material, we generate a spatially varying dielectric environment that modifies both the band edges and exciton binding energy. This results in an effective lateral potential well capable of supporting localized exciton states—enabling deterministic, lithography-compatible single-photon emission without relying on strain, defects, or etching of the 2D layer. Using numerically calculated interaction potentials and solutions to the 2D excitonic Schrödinger equation, we show that the confinement strength, energy levels, and spatial extent of excitons can be precisely tuned by adjusting the pillar's geometry and dielectric properties. We quantify the resulting confinement through binding energies, wavefunction profiles, and a spatial confinement parameter, and demonstrate that confinement depths exceeding 100 meV are achievable. These results offer a practical and scalable route for integrating 2D-based quantum emitters into photonic platforms, paving the way for next-generation quantum optoelectronic devices.

## I. INTRODUCTION.

Single-photon emitters (SPEs) are foundational components in emerging quantum technologies [1,2], enabling secure quantum communication [3,4], quantum key distribution [5], and scalable quantum photonic networks [6,7]. To realize large-scale, practical quantum systems, it is essential to integrate such quantum light sources on-chip in a scalable, robust, and CMOS-compatible platform [8,9]. The integration of SPEs with conventional semiconductor processing technology would allow seamless coupling to photonic circuits [8], paving the way for high-density quantum photonic devices operating at ambient conditions [2,6].

Two-dimensional (2D) materials, such as transition metal dichalcogenides (TMDs), have emerged as promising candidates for room-temperature single-photon emission due to their strong excitonic effects and atomically thin geometry [10–17]. In particular, monolayer TMDs exhibit tightly bound excitons with large binding energies [18,19] and are highly sensitive to their surrounding dielectric environment [20–24], offering a versatile platform for photonic engineering [25]. However, generating localized and deterministic quantum emitters in 2D materials remains a central challenge [26–30]. Most demonstrations rely on strain engineering [28,31,32], defects [33,34], nanobubble [35], or nanoantennas [36], which are either non-deterministic, or non-repudiable, or difficult to integrate into standard fabrication workflows.

An alternative and scalable route to exciton confinement in two-dimensional semiconductors is provided by local dielectric engineering, which enables controlled modification of Coulomb interactions within the 2D layer [22,37,38]. Because excitonic binding in these materials is strongly influenced by the surrounding dielectric environment, spatial structuring of nearby dielectrics offers a versatile means to tailor the exciton energy landscape and induce lateral confinement [23,39]. Experimental and theoretical studies have demonstrated that dielectric modulation can lead to exciton localization in two-dimensional systems, highlighting the potential of this approach for engineered quantum states [40].

In this work, we present a fully numerical framework that self-consistently evaluates the spatially varying excitonic potential and self-energy in realistic dielectric configurations. This approach enables a quantitative description of exciton confinement and spatial extent, and allows direct identification of regimes where deterministic single-exciton occupancy emerges as the characteristic size of the dielectric feature approaches the exciton radius. Based on this framework, we propose and analyze a platform for exciton confinement employing a high-dielectric-constant (high-κ) nanopillar positioned either on top of or beneath the 2D semiconductor, offering a scalable pathway toward deterministic quantum light sources.

The presence of a high-$k$ dielectric pillar introduces a localized enhancement of dielectric screening, which modifies both the exciton binding energy and the single-particle band gap beneath the pillar. As a result, the exciton transition energy is lowered locally, forming an effective lateral potential well that can trap the exciton in a confined state. If the confinement is sufficiently strong, it supports discrete energy levels analogous to those in semiconductor quantum dots, enabling single-photon emission under optical excitation [41]. Notably, this dielectric-based approach does not require structural modification of the 2D material, allowing for scalable and CMOS-compatible fabrication with deterministic spatial control over emitter placement and energy [8,42].

Recent studies have shown that increasing dielectric screening in two-dimensional materials induces a simultaneous reduction in both exciton binding energy ($E_b$) and the quasiparticle band gap ($E_g$), a phenomenon commonly referred to as band gap renormalization (BGR) [43]. However, these two energy shifts do not scale equally. Multiple reports consistently show that the reduction in the band gap typically exceeds that of the exciton binding energy, with a typical ratio: $\frac{\Delta E_g}{\Delta E_b} \sim 2$ observed across various materials and dielectric environments [44–48]. This implies that for every 1 meV decrease in exciton binding energy, the band gap is reduced by approximately 2 meV. Consequently, the net reduction in exciton transition energy ($\Delta E_x = \Delta E_g - \Delta E_b$) can be approximated by: $\Delta E_x \sim \Delta E_b$ which highlights that changes in the binding energy are a reliable proxy for estimating local exciton energy shifts. Therefore, maximizing the reduction in exciton binding energy, by increasing $\varepsilon_{pillar}$, the dielectric contrast of the pillar, results in an even greater decrease in the band gap, effectively reinforcing the exciton confinement potential.

*Contact author: ilya.goykhman@mail.huji.ac.il

To allow only one 2D exciton to be presence in the confinement, the $r_{pillar}$, the radius of the pillar should be taken into consideration carefully. pillar with radius larger than, $r_{ex}$, the 2D exciton radius, will allow many excitons to be presence in the 2D material beneath the pillar area. Smaller pillar radius blocks the exciton to be presence in the 2D material beneath the pillar area. Hence, the radius of the pillar should be as the size of the radius of the 2D exciton $r_{pillar} \sim r_{ex}$.

Since the exciton radius increases with stronger dielectric screening [49–51], the radius of the high-$k$ pillar must also be increased to maintain effective spatial confinement. As the pillar radius increases, the exciton wavefunction becomes more extended, up to the point where the pillar radius exceeds the exciton radius. Beyond this threshold, further increasing the pillar size does not significantly enhance confinement of a single exciton but instead creates a potential landscape capable of accommodating multiple excitons within the same region. Therefore, the dielectric constant and the radius of the pillar must be carefully optimized in relation to each other to achieve strong confinement while avoiding unintentional multi-exciton occupancy.

Figure 1(a) illustrates the main device geometry considered when the pillar high-k dielectric material is placed directly on top of the 2D layer. Another approach is to embed the pillar within the SiO$_2$ substrate as shown in Figure 1(b). The main different between the two is that in the latter, the transfer of the 2d material is done after the lithography part of the pillar. In both cases, the dielectric discontinuity modifies the electrostatic interaction between charged carriers, resulting in an attractive potential well for excitons located within the vicinity of the pillar. This spatial modulation in dielectric screening is translated into a lateral confinement potential, as depicted schematically in Figure 1(c). The confinement acts primarily through reduction of the exciton binding energy beneath the pillar, effectively creating a potential well.

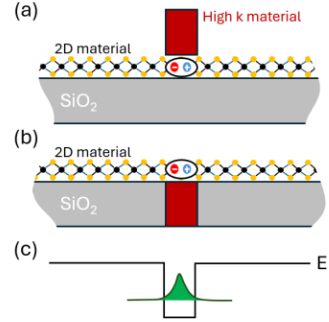

Figure 1. Device schematics showing a high-k dielectric (a) on top of the 2D material and (b) embedded in the SiO$_2$ substrate. (c) Corresponding schematic confinement potential and wavefunction.

In the following sections, we present a comprehensive theoretical investigation of this dielectric confinement scheme. We calculate the screened interaction potential in the presence of a high-k pillar, solve for the bound exciton states, and characterize their energies, spatial profiles, and confinement strength as a function of the dielectric and geometric parameters of the pillar. Our results demonstrate a robust and tunable mechanism for generating localized exciton states in 2D materials, providing a promising route toward scalable, on-chip quantum light sources.

To achieve spatial confinement of excitons in a two-dimensional (2D) semiconductor, we exploit dielectric engineering by locally modifying the surrounding dielectric environment using a high-permittivity (high-$k$) nanostructure. Specifically, we place a dielectric pillar above (or embedded within) the substrate beneath the 2D material, resulting in a spatially varying screening environment that modifies the exciton binding potential. This approach provides an effective method to create a quantum dot-like confinement potential without introducing disorder or altering the 2D material itself.

## II. RESULTS

To accurately quantify the electrostatic environment experienced by excitons in two-dimensional (2D) materials, we numerically calculate the interaction potential of a point charge embedded in a 2D layer situated on a SiO$_2$ substrate. The computed potential is compared to two widely used analytical models: the bare Coulomb potential and the Rytova–Keldysh (RK) potential. Figure 2(a) presents the comparison between the numerically obtained potential (black curve), the Coulomb potential (blue), and the RK potential (red), while Figure 2(b) shows the corresponding relative errors.

*Contact author: ilya.goykhman@mail.huji.ac.il

Consistent with previous studies [52], the interaction potential can be categorized into three characteristic regimes based on the in-plane distance r from the charge relative to the 2D material thickness d. In the short-range regime (r≪d), the potential is well approximated by the Coulomb form due to dominant in-plane screening. In the long-range regime (r≫d), the potential transitions to the RK form, which accounts for dielectric screening from the surrounding environment. In the intermediate regime (r∼d), neither the Coulomb nor RK models capture the full behavior of the potential accurately.

To address this, we employ a numerical solution of the Poisson equation in the multilayer dielectric geometry, enabling an accurate description of the interaction potential across all spatial scales. This comprehensive approach ensures reliable modeling of the exciton's electrostatic environment, particularly in the transitional regime where approximate analytical models fail.

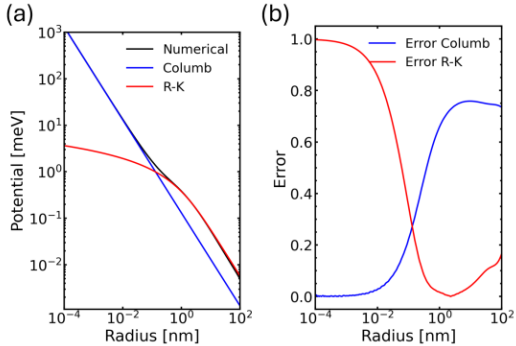

**Figure 2.** (a) Numerically calculated potential of a point charge in a 2D material on a SiO$_2$ substrate, compared with the Coulomb and Rytova-Keldysh (R-K) models. (b) Relative errors of the Coulomb and R-K approximations with respect to the numerical result.

To explore how the geometry and dielectric properties of the pillar influence the spatial interaction potential, we numerically calculated the electrostatic potential of a point charge in a 2D material as a function of the dielectric constant (ε) and radius (r) of the high-*k* dielectric pillar. Figure 3(a) shows the variation of the potential profile with increasing dielectric constant. As ε increases, the screening becomes stronger, resulting in a reduction of the potential beneath the pillar by more than an order of magnitude. Figure 3(b) presents the effect of varying the pillar radius. Increasing the radius leads not only to a greater suppression of the potential at the center but also to an extended spatial region over which the potential is reduced. These observations indicate that larger and more polarizable dielectric pillars create deeper and broader attractive potentials for excitons, facilitating stronger spatial localization. By carefully optimizing both the dielectric constant and the radius of the pillar, the confinement potential can be engineered to form an effective trap for 2D excitons.

*Contact author: ilya.goykhman@mail.huji.ac.il

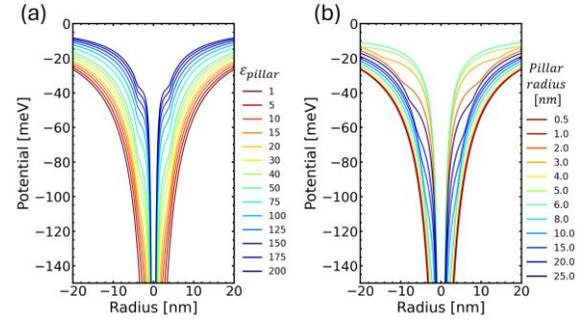

**Figure 3.** Numerical potential of a point charge in a 2D material on a SiO$_2$ substrate with a high-k pillar placed on top, calculated for varying (a) dielectric constants with pillar radii of 10 nm and for varying (b) pillar radii with dielectric constant of 20.

Next, we numerically solve the 2D Schrödinger equation to obtain the bound exciton states under various dielectric configurations. Figure 4 presents the squared wavefunctions $|\psi(x)|^2$ and their corresponding energy levels for the four lowest s-states (1s to 4s), under increasing values of dielectric constant ε and pillar radius r. As expected, higher ε and larger r lead to reduced exciton binding energies and broader wavefunctions, corresponding to a larger exciton radius. As discussed previously, increasing the spatial dielectric function not only enhances the reduction in the interaction potential but also expands the exciton's spatial extent. To examine this effect, we plot the exciton wavefunctions and their corresponding energies for several quantum states under different configurations.

In the absence of the pillar (ε=1), the exciton experiences strong Coulomb attraction, resulting in a deeply bound and highly localized 1s state, as shown in Figure 4(a). Introducing a dielectric pillar with ε=20 and r=20 nm significantly reduces the binding energy and increases the spatial extent of the wavefunctions, as shown in Figure 4(b). Increasing the dielectric constant further to ε=100, while reducing the pillar radius to r=4 nm, results in even lower binding energies and broader exciton states, as shown in Figure 4(c). Finally, when both parameters are increased (ε=100, r=8nm), the binding energy drops substantially, and the exciton wavefunctions become even more extended, as illustrated in Figure 4(d). These results demonstrate the strong sensitivity of the exciton's spatial localization and energy spectrum to the local dielectric environment, enabling tunable control over exciton confinement through nanoscale engineering of dielectric structures.

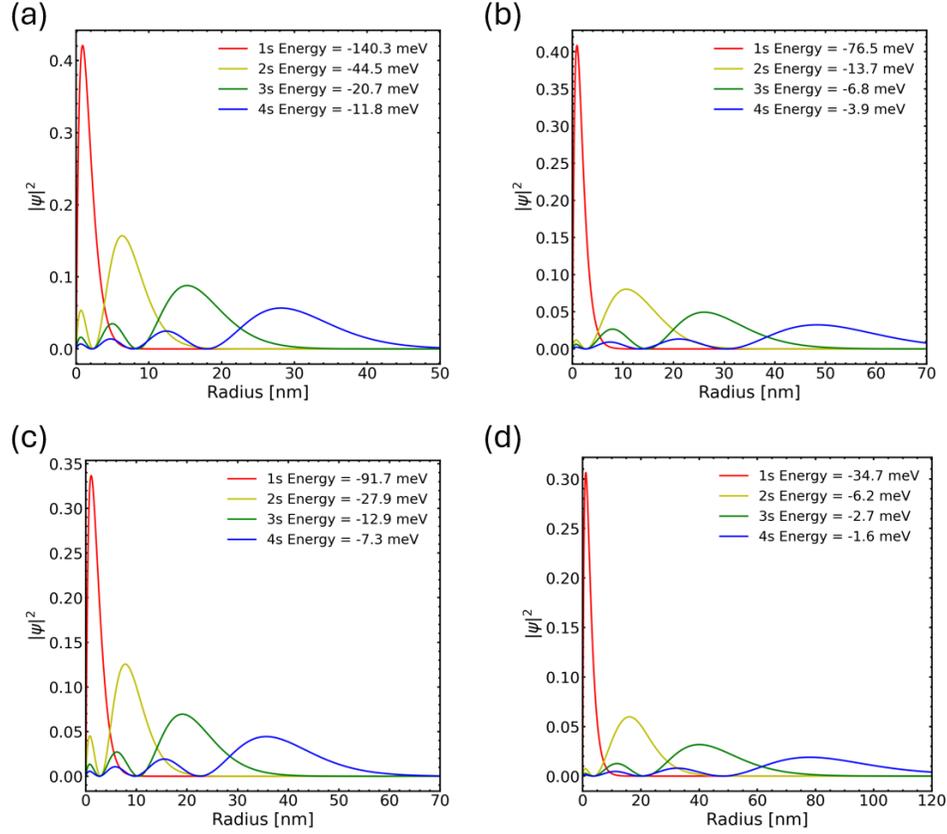

**Figure 4.** 2D exciton wavefunctions and corresponding energy levels for various dielectric environments. (a) pillar with ε = 20 and radius = 4 nm, (b) pillar with ε = 20 and radius = 20 nm, (c) ε = 50 and radius = 4 nm, and (d) ε = 50 and radius = 20 nm.

To quantify spatial confinement, we define a confinement parameter $\Gamma$, which represents the fraction of the exciton wavefunction localized beneath the pillar. It is given by:

$$\Gamma = \frac{\int_0^{r_{pillar}} |\psi(x)|^2 dx}{\int_0^\infty |\psi(x)|^2 dx}$$

Figures 5(a) and 5(b) present $\Gamma$ as a function of the dielectric constant and the pillar radius for the 1s and 2s exciton states, respectively. For the 1s state, when the pillar radius exceeds 4 nm, the confinement parameter remains above 0.8 and shows relatively weak dependence on the dielectric constant. In contrast, the 2s state displays strong sensitivity to both radius and dielectric constant: the confinement decreases sharply with increasing ε, indicating that the 2s state is more easily delocalized beyond the pillar. This behavior implies that while the 1s state remains well-confined beneath the pillar, higher-order states such as 2s can be excluded from the pillar region under strong screening. Figure 5(c) directly compares the two states for a fixed pillar radius of 4 nm. While the confinement of the 1s state remains above 0.8 for all values of ε, the 2s confinement remains below 0.3, confirming that only the 1s state is significantly localized beneath the pillar in this configuration. These results emphasize the role of both pillar geometry and dielectric contrast in selectively confining specific excitonic states, offering a route toward tailored quantum-level engineering of 2D excitons.

To further explore the distinct confinement behavior of the 1s and 2s exciton states, we analyze their spatial extent by calculating the expectation value of the exciton radius, ⟨r⟩, obtained from the radial probability distribution. Figures 6(a) and 6(b) show the expectation radius as a function of the pillar radius and dielectric constant for the 1s and 2s states, respectively. For the 1s state, the exciton radius increases rapidly for small pillar radii and then saturates at approximately 4 nm, with a moderate dependence on the dielectric constant. In contrast, the 2s state exhibits a much stronger dependence on both dielectric constant and pillar radius, with the expectation radius exceeding 20 nm under strong screening conditions (large ε and r). Figure 6(c) directly compares the two states for a fixed pillar radius of 4 nm. While the exciton radius of the 1s state remains around 1-2 nm for all values of ε, the 2s exciton radius rapidity increases from 5nm to 9 nm.

*Contact author: ilya.goykhman@mail.huji.ac.il

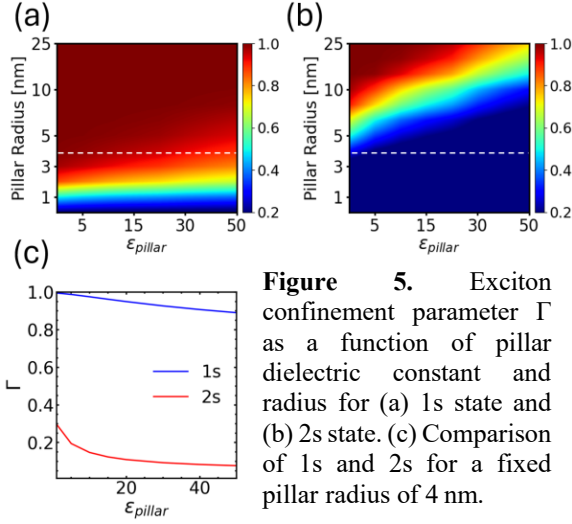

**Figure 5.** Exciton confinement parameter Γ as a function of pillar dielectric constant and radius for (a) 1s state and (b) 2s state. (c) Comparison of 1s and 2s for a fixed pillar radius of 4 nm.

These results are consistent with the confinement behavior discussed previously and demonstrate that the spatial extent of the exciton can be precisely engineered through local modifications of the dielectric environment. The 1s state remains tightly bound and localized within the pillar region, whereas higher-order states such as 2s become increasingly delocalized as the dielectric contrast and confinement dimensions grow.

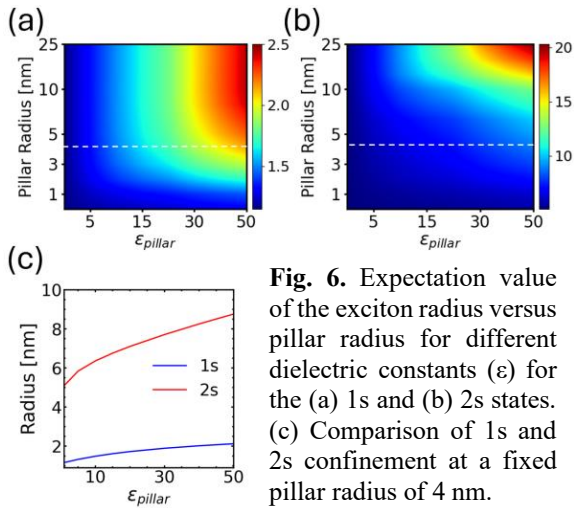

**Fig. 6.** Expectation value of the exciton radius versus pillar radius for different dielectric constants (ε) for the (a) 1s and (b) 2s states. (c) Comparison of 1s and 2s confinement at a fixed pillar radius of 4 nm.

To complete the analysis, we examine the binding energies (self-energies) of the 1s and 2s exciton states under varying dielectric environments. Figure 7 presents the calculated binding energies of both states as functions of the pillar radius and dielectric constant. As expected, the binding energies increase (i.e., become less negative) with increasing ε and pillar radius, reflecting enhanced dielectric screening and a corresponding reduction in Coulomb attraction. For the 1s state, the energy saturates for pillar radii larger than approximately 4 nm, indicating that the exciton becomes insensitive to further increases in the confinement size. In contrast, the 2s state continues to show a strong dependence on both ε and radius, with the binding energy increasing consistently across the entire parameter range. These results highlight a key distinction between ground and excited exciton states: while the 1s state reaches a saturation regime due to its compact nature, higher-order states like 2s remain more responsive to changes in the dielectric landscape, providing further opportunities for selective exciton manipulation through nanoscale dielectric engineering.

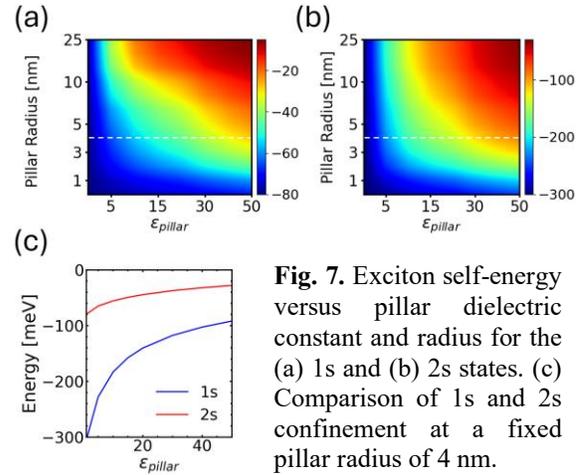

**Fig. 7.** Exciton self-energy versus pillar dielectric constant and radius for the (a) 1s and (b) 2s states. (c) Comparison of 1s and 2s confinement at a fixed pillar radius of 4 nm.

To quantify the effective confinement depth created by the presence of the dielectric pillar, we calculated $\Delta E_b$, the difference in exciton binding energy between regions beneath and outside the pillar. Figure 8(a) presents this binding energy difference as a function of the pillar dielectric constant and radius. The results show that $\Delta E_b$ increases with the dielectric constant and saturates for pillar radii larger than approximately 4 nm, indicating that the confinement strength becomes limited by screening rather than geometry beyond this point.

Figures 8(b) and 8(c) illustrate the discrete energy levels of the exciton outside and beneath the pillar, respectively, for the case of ε=20 and r=4 nm. In this configuration, the binding energy outside the pillar is

*Contact author: ilya.goykhman@mail.huji.ac.il

$E_b = 305\ meV$, while beneath the pillar it is $E_b = 140\ meV$. This results in a confinement depth of approximately 165 meV, which is sufficient to localize the exciton within a single quantum level under low-temperature conditions.

These results demonstrate that spatial modulation of the dielectric environment via a high-ε pillar introduces a robust exciton trapping potential, offering a deterministic and lithography-compatible platform for localized single-photon emitters in 2D semiconductors.

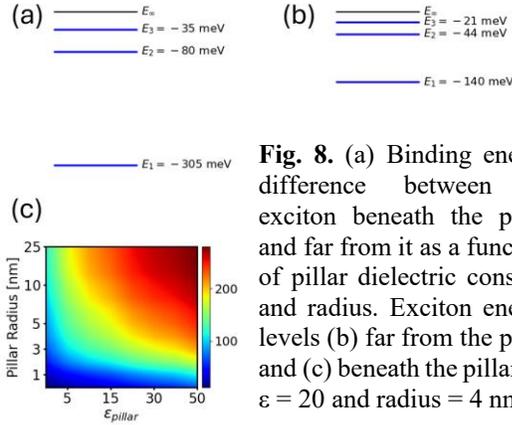

**Fig. 8.** (a) Binding energy difference between an exciton beneath the pillar and far from it as a function of pillar dielectric constant and radius. Exciton energy levels (b) far from the pillar and (c) beneath the pillar for ε = 20 and radius = 4 nm.

Finally, to capture the lateral behavior, we solved the full two-dimensional exciton potential and calculated the corresponding energy levels as a function of the pillar position, thereby illustrating the influence of local dielectric modulation on exciton confinement in a monolayer TMD supported by a structured substrate. Figure 9(a) shows the schematic of the numerical model, where a high-dielectric pillar embedded in the upper substrate is laterally displaced along the x-axis beneath the TMD layer. This displacement alters the local dielectric screening, leading to a spatially varying potential landscape that strongly depends on the pillar position. The resulting potential profiles, shown in Figure 9(b), exhibit a pronounced well centered above the pillar that becomes shallower and asymmetric as the pillar shifts away from the exciton center, indicating a gradual reduction in confinement strength.

The simulated exciton wavefunctions in Figure 9(c) show that the 1s state remains highly localized within the pillar region, while the 2s state extends far beyond it, consistent with its weaker binding and reduced sensitivity to the dielectric perturbation. This comparison reveals that the energy modulation induced by the pillar primarily affects the 1s state, as higher-order states experience much weaker confinement and remain nearly delocalized. Notably, the lateral extent of the 1s exciton is comparable to the pillar diameter, suggesting that only a single exciton can be localized beneath a single pillar at a time—an essential condition for achieving deterministic single-photon emission.

The corresponding exciton energy levels in Figure 9(d) further confirm this behavior: as the pillar shifts laterally, the 1s binding energy shows a strong position-dependent modulation, while excited-state energies remain relatively unaffected. The dashed lines mark the pillar boundaries, highlighting the direct correspondence between the nanoscale dielectric geometry and the spatially confined excitonic potential.

## III. DISCUSSION

This work demonstrates that dielectric engineering via a high-permittivity nanopillar enables robust and tunable exciton confinement in two-dimensional semiconductors. By spatially modulating the surrounding dielectric environment, both the exciton binding energy and the quasiparticle band gap are reduced. However, the reduction in the band gap consistently outweighs the binding energy suppression, resulting in a net decrease in exciton transition energy beneath the pillar. This dielectric-induced energy modulation gives rise to an effective lateral potential well capable of trapping excitons.

Our simulations reveal that the depth and width of the confinement potential are strongly influenced by two parameters: the dielectric constant (ε) and the radius of the pillar. As shown in Figure 3, increasing either parameter enhances the reduction in the Coulomb potential, thereby strengthening exciton confinement. However, beyond a certain threshold (e.g., pillar radius > 5 nm), the potential profile begins to saturate, suggesting diminishing returns with further geometric enlargement.

By numerically solving the 2D Schrödinger equation for various configurations, we show that both the exciton binding energy and wavefunction extent are highly sensitive to the spatial dielectric profile. The ground-state (1s) exciton remains localized beneath the pillar across a wide parameter range, whereas higher-order states (e.g., 2s) are more weakly bound and easily delocalized. The confinement parameter Γ confirms that only the 1s state remains spatially confined when the dielectric contrast is strong, while 2s states extend beyond the pillar boundary. This selective confinement opens pathways for controlling individual quantum states using passive, patterned dielectric features.

Further insight is provided by analyzing the expectation value of the exciton radius, which increases with both ε and the pillar radius, particularly for the excited states (Figure 6). The 2s expectation radius grows significantly with increasing pillar size and dielectric constant, exceeding 30 nm under strong screening conditions. This behavior indicates that while the 1s exciton remains tightly confined beneath the pillar, higher-order states become progressively more delocalized. As shown in Figure 7, the 1s binding energy saturates beyond moderate pillar sizes, whereas the 2s state remains highly sensitive to both $\varepsilon$ and pillar radius.

*Contact author: ilya.goykhman@mail.huji.ac.il

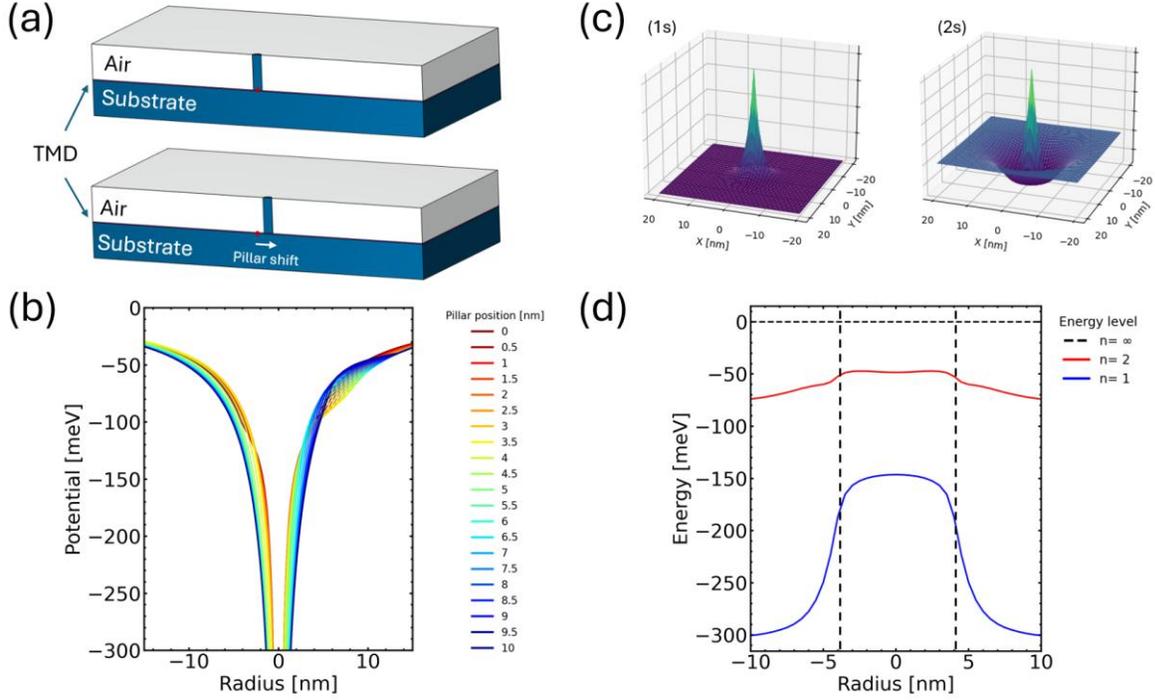

**Figure 9.** (a) Schematic illustration of the numerical model showing a dielectric pillar embedded in the substrate and its lateral displacement along the x-axis beneath the TMD layer. (b) Calculated exciton potential profiles along the x-direction for various pillar positions, demonstrating the evolution of the confinement potential with increasing lateral shift. (c) Simulated 2D exciton wavefunctions corresponding to the 1s and 2s states for a pillar of radius 4 nm and dielectric constant $\varepsilon = 20$ located at the origin. (d) Calculated exciton energy levels as a function of lateral position, with dashed lines indicating the pillar boundaries.

Figure 8 demonstrated that the exciton energy difference between regions inside and outside the pillar can exceed 100 meV, providing a substantial confinement potential. Extending this analysis, Figure 9 highlights the lateral behavior of the confined exciton. The energy modulation primarily affects the 1s state, whose spatial extent closely matches the pillar diameter. This correspondence indicates that each pillar can host a single confined exciton, creating a deterministic, site-controlled quantum emitter. In contrast, excited states remain weakly perturbed and spatially extended, reinforcing that the dielectric confinement acts as a single-level potential well. Together, these results demonstrate that dielectric-defined energy landscapes can produce discrete excitonic states without structural etching or mechanical strain, paving the way for scalable and CMOS-compatible exciton confinement architectures.

### IV. CONCLUSION

We have proposed and theoretically demonstrated a scalable, deterministic approach for exciton confinement in two-dimensional semiconductors using dielectric nanopillars. By engineering the local dielectric environment, we achieve simultaneous modulation of the exciton binding energy and band edge energies, resulting in spatial confinement of the exciton transition energy. Numerical simulations confirm that the 1s exciton can be strongly localized beneath a high-$\varepsilon$ pillar, while higher-order states remain delocalized, enabling selective quantum state control.

The strength and extent of confinement are tunable via the dielectric constant and geometry of the pillar. Importantly, our results reveal that band-gap renormalization dominates over binding energy reduction, producing a net redshift of the exciton transition energy that defines an effective lateral potential well. The depth of this well can exceed 140 meV, sufficient for single-level confinement under cryogenic and potentially even room-temperature conditions.

*Contact author: ilya.goykhman@mail.huji.ac.il

This dielectric confinement strategy offers a promising route to quantum-dot-like exciton localization without etching, strain, or chemical modification, and is fully compatible with CMOS fabrication processes. As such, it provides a robust and scalable platform for on-chip single-photon emitters and other integrated quantum photonic devices based on 2D materials.

The Discussion section may be divided into subheadings and may be combined with the Results section.

## METHODS

The interaction potential of a point charge in the presence of a dielectric pillar was calculated numerically by solving the Poisson equation in a multilayer dielectric environment using finite-element methods. The 2D material was modelled as a thin sheet with dielectric constant of 15 and thickness of 0.7 nm located at the interface between vacuum and the $SiO_2$ substrate. The dielectric pillar was introduced as a cylindrical inclusion of radius $r$ and permittivity $\varepsilon$, on top of the substrate. The point charge was defined using delta function with strong of e charge.

The exciton problem was treated in the effective mass approximation. The electron-hole interaction potential obtained from the electrostatic simulations was used as input for the 2D Schrödinger equation:

$$\left[-\frac{\hbar^2}{2\mu}\nabla^2 + V(r)\right]\psi(r) = E\psi(r)$$

where $\mu=0.16m_e$ is the reduced exciton mass and $V(r)$ is the screened interaction potential. The equation was solved numerically using finite-difference discretization and diagonalization of the Hamiltonian matrix.

The confinement parameter $\Gamma$ was computed as the integral of $|\psi(x)|^2$ over the spatial region beneath the pillar. Throughout this work, we extract the "classical" exciton radius as the expectation value of the distance, ⟨r⟩, computed directly from the numerically obtained 2D wavefunction.

The data processing was performed using in house program in Python.

All The numerical simulation, coding and data can be given by reasonable request.

## ACKNOWLEDGMENTS

The authors thank Technion for providing computational resources.

*Contact author: ilya.goykhman@mail.huji.ac.il

*Contact author: ilya.goykhman@mail.huji.ac.il

*Contact author: ilya.goykhman@mail.huji.ac.il